\documentclass[aps,prb,superscriptaddress,reprint]{revtex4-1}%
\usepackage{epsfig,pslatex,latexsym,times,amssymb,amsmath,graphicx}
\usepackage{bm, float, lipsum}
\usepackage{amsmath}
\usepackage{amsfonts}
\usepackage{amssymb}
\usepackage{mathrsfs, notes2bib}
\usepackage{color}
\usepackage{hyperref}
\usepackage{graphicx}%
\providecommand{\U}[1]{\protect\rule{.1in}{.1in}}
\begin{document}
\author{N. J. Harmon}
\email{nicholas-harmon@uiowa.edu} 
\author{M. E. Flatt\'e}
\affiliation{Department of Physics and Astronomy and Optical Science and Technology Center, University of Iowa, Iowa City, Iowa
52242, USA}
\date{\today}
\title{Spin Relaxation in Materials Lacking Coherent Charge Transport}
\begin{abstract}
We describe a broadly-applicable theory of spin relaxation in materials with incoherent charge transport;  examples include amorphous inorganic semiconductors, organic semiconductors, quantum dot arrays, and systems displaying trap-controlled transport or transport within an impurity band.
The theory can incorporate many different relaxation mechanisms, so long as electron-electron correlations can be neglected. 
We focus primarily on spin relaxation caused by spin-orbit effects, which manifest through inhomogeneities in the $g$-factor and non-spin-conserving carrier hops, scattering, trapping, or detrapping.
Analytic and numerical results from the theory are compared in various regimes with Monte Carlo simulations.
Our results should assist in evaluating the  suitability of various disordered materials for spintronic devices.
\end{abstract}
\maketitle

\section{Introduction}

Spin relaxation associated with the band transport of electrons in nonmagnetic materials exhibits a variety of regimes and mechanisms, depending on lattice symmetries, the ratio of momentum scattering and spin-orbit-interaction times, and the presence of nuclear spin interactions\cite{Yafet1963,Meier1984,Ziese2001,Awschalom2002,Samarth2004}. When the electrons are localized they relax via different mechanisms, such as spin-spin interactions\cite{Abragam1961,Slichter1996}. Considerably less attention has been directed towards spin relaxation in systems where charge transport occurs through incoherent motion.
These largely focused on organic (non-crystalline) semiconductors\cite{Dediu2002, Xiong2004, Bobbert2009, Nguyen2010, Baker2012b, Rybicki2012, Harmon2013a}, due to their small spin-orbit interaction, affordability, and  large room temperature spin-dependent effects\cite{Vardeny2010}. Spin transport in disordered crystalline semiconductors has been used as a diagnostic tool for very small numbers of defects in semiconductor junctions using electrically-detected magnetic resonance \cite{Cochrane2012},
but has not drawn the same attention to fundamental mechanisms in macroscopic materials as these other systems. 
Aspects of the spin transport problem when charge transport is incoherent also have been studied  in systems demonstrating impurity band transport\cite{Tamborenea2007, Intronati2012}, arrays of quantum dots\cite{Zinovieva2010, Zinovieva2014}, and amorphous inorganic semiconductors\cite{VogetGrote1976, Movaghar1977, Movaghar1978}.
A fuller understanding of spin relaxation in disordered semiconductors would help clarify the behavior of spintronic devices based on these materials, such as spin valves.\cite{Xiong2004, Pramanik2007, Grunewald2011, Drew2009, Dediu2010} The influence of spin relaxation on light emitting diodes and solar cells has recently become a focus of considerable interest, due to results showing changes in (sometimes improving substantially) device performance when spin relaxation is increased in the materials.\cite{Nguyen2012, Zhang2012, Wang2014} Although the investigation of spin relaxation in amorphous inorganic semiconductors has received considerably less attention than that of organic semiconductors, similar effects can be expected in such materials

In this article we generalize our previous work\cite{Harmon2013a} with organic semiconductors to describe spin relaxation in a broader range of regimes of incoherent charge transport, focusing on amorphous semiconductors such as silicon (a-Si) and germanium (a-Ge) to showcase our results.
This is done by explicit calculations of spin lifetimes and coherence times using continuous-time random walk (CTRW) theory \cite{Montroll1965, Scher1973} as well as with Monte-Carlo simulations. Our theory allows us to make precise predictions. Amorphous semiconductors are attractive theoretically since they exhibit ``dispersive transport" which possesses features explainable by disordered transport theories.\cite{Scher1975, Pfister1978, Jakobs1993} Analysis of transport is murkier for organic semiconductors due to their supposed Gaussian density of states.\cite{Hartenstein1996}

In the theory presented herein, the pivotal transport characteristic is the \emph{wait-time distribution} (WTD).
This quantity, elemental to CTRW theory, describes the probability density function for wait-times between transport-related events.
Most often these events are hops between localizing centers but could also signify trapping and trap-release times.
The WTD very much depends on the system and regime under consideration; because the wait-times typically  depend on the energy depth of a charge's inhabitance, the energy density of states plays an important role in determining the WTD.
Determinations of the density of states and the WTD are vital quantities to be ascertained for the various disordered systems.
\begin{figure}[ptbh]
 \begin{centering}
        \includegraphics[scale = 0.29,trim = 70 320 100 160, angle = -0,clip]{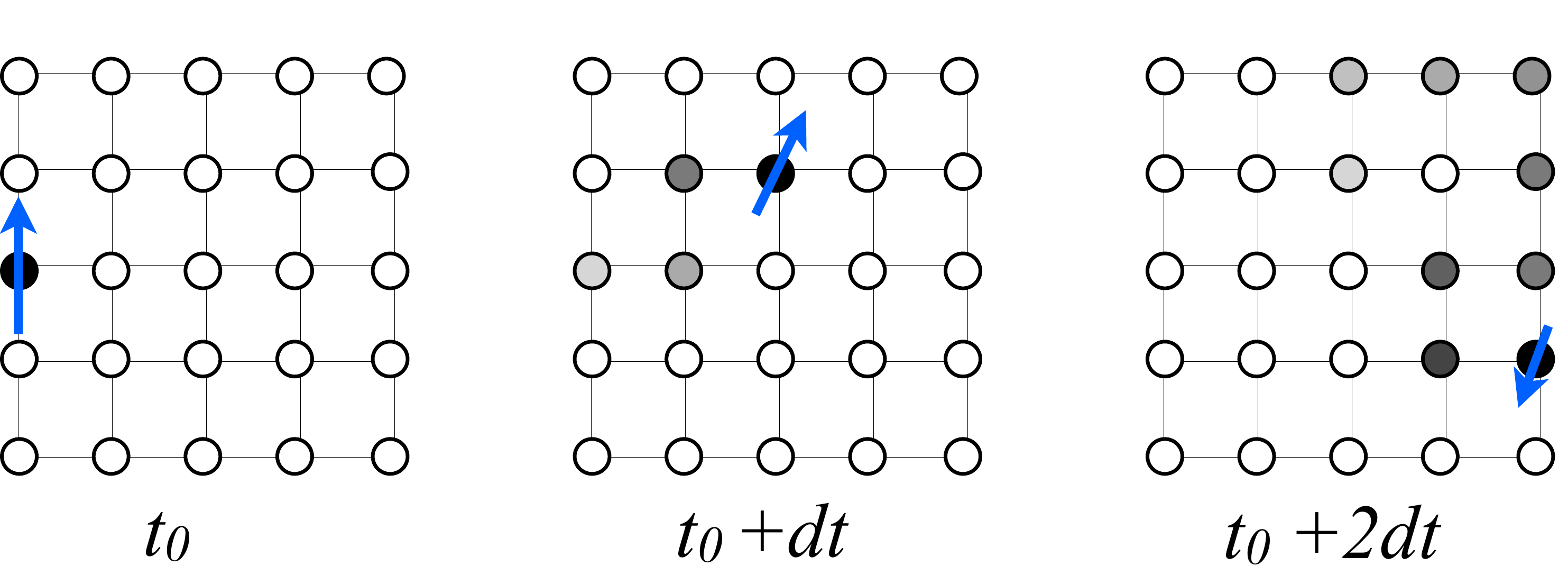}
        \caption[]
{(Color online) Example for the spin evolution of a randomly walking charge. At $t_0$, the spin is initialized in a specific direction. At some time interval, $dt$ later, the charge has hopped several times and is now at a new location (its history is denoted by the shading of the sites). Due to local fields during the walk, the spin has undergone rotations. At time $t_0 + 2 dt$ the charge has moved further, and the spin has rotated further towards the point of a spin  flip.}\label{fig:rw}
        \end{centering}
\end{figure}

Incoherent charge transport is treated as a random walk between the various states present in the system. 
As a carrier randomly walks, we keep track of its classical spin vector given some set of spin interactions which we model as local magnetic fields. Some of these fields are exerted on the spin in between steps; such fields result in a spin rotation given by $\hat{\bm{R}}_s$ ($s$ for stationary). The spin may be influenced by other local fields during the stepping process; they rotate by $\hat{\bm{R}}_h$ ($h$ for hop).
We assume the \emph{strong collision approximation}\cite{Hayano1979} for the spin random walk which entails that the local fields change instantaneously and are completely uncorrelated from step to step. Wait-times at any position or instant are also uncorrelated from one another. We consider correlated steps in our simulations.
Figure \ref{fig:rw} captures the evolution of a spin while it randomly walks.
The beauty of the CTRW theory is that a spin ensemble's various random rotations that are incurred from its random walk can be summed exactly to determine the spin polarization.

The output of our calculations are spin polarization functions of time; depending on the system at hand we can sometimes analytically determine the longitudinal (transverse) spin relaxation (decoherence) time which we denote as $T_1$ ($T_2$), respectively. In more complicated cases, $T_1$ and $T_2$ can be extracted from numerical fits to the spin polarization functions. In some cases of dispersive transport, spin losses are algebraic in time and a characteristic timescale is ill-defined.

We concentrate on the two mechanisms reported as dominant in a:Si: inhomogeneous $g^*$ ($\delta g$) spin dephasing/decoherence and spin-orbit coupling (SOC) induced spin relaxation. 
The $\delta g$ mechanism occurs due to variations in $g^*$ felt by spins in magnetic fields.
In the static limit, this mechanism is a reversible dephasing process; however the occurrence of hopping leads to irreversible decoherence. 
The SOC mechanism results from impure spin states where the amount of spin admixture is tied to the magnitude of the SOC.
In such a case, the spin-flip matrix elements are non-zero for scattering from spin independent potentials.

We point out the organization of the article by summarizing the primary results:
in Section \ref{section:II}, the central results of the CTRW theory of spin relaxation are derived. A compact expression for the spin polarization is obtained.
Examples of the theory are given in Sections \ref{section:III} and \ref{section:IV}, using the $\delta g$ and SOC mechanisms.
We discover that strong disorder dramatically alters SOC relaxation such that the  decay is algebraic instead of exponential.
The significance of correlations between hopping events is shown to be small for $\delta g$  but large for SOC. 
Section \ref{section:V} provides a generalization of the theory to situations where transport is governed by crossings between two types of transport states. An example of such a system is trapping and detrapping between extended and localized states.
In Section \ref{section:VI}, we apply the theory to amorphous inorganic semiconductors (a:Si specifically) and compare to available experiments. The position of the Fermi level is important in determining the transport, as well as spin lifetime, regime in these systems. Undoped samples exhibit hopping within strongly localized states. Since hopping is promoted by temperature, low temperature polarization loss is governed by the static limit of the $\delta g$ mechanism. 
At higher temperatures (and faster hopping), the SOC mechanism is predominant. 
Doped systems behave differently since the Fermi level is raised near the mobility edge. At this level, both localized and extended states play a role in the spin relaxation, which is well described by our theory.

\section{Theory}\label{section:II} 

\subsection{Continuous Time Random Walk Theory of Spin Relaxation}

The unit vector, $\bm{S}$, is a classical spin which in an arbitrary static field, $\bm{\omega} = \mu_B \hat{\bm{g}} \bold{B}/  \hbar$, is described by the following evolution:
\begin{equation}\label{eq:spinDiffEq}
\frac{d\bm{S}(t)}{dt} = \bm{\omega} \times \bm{S}(t)  - \Gamma \bm{S}(t) = \bm{\Omega} \cdot \bm{S}(t) - \Gamma \bm{S}(t),
\end{equation}
where $\bm{\Omega}$ is the skew-symmetric matrix
\begin{eqnarray}
\bm{\Omega} &=& \omega  \hat{\bm{\Omega}} = 
\left( {\begin{array}{ccc}
 0 & -\omega_z  & \omega_y  \\
  \omega_z & 0  & -\omega_x \\
 -\omega_y & \omega_x  & 0  \\
 \end{array} } \right) {}\nonumber\\
 &\equiv&{} \omega
 \left( {\begin{array}{ccc}
 0 & -\cos\theta  & \sin\theta \sin\phi  \\
  \cos\theta & 0  & -\sin\theta \cos\phi \\
 -\sin\theta \sin\phi& \sin\theta \cos\phi  & 0  \\
 \end{array} } \right)
\end{eqnarray}
and $\hat{\bm{g}}$ is in general a tensor of the g-factors which can be expressed as a 3$\times$3 matrix;
the magnitude of the precession frequency is 
$\omega = \mu_B|\hat{\bm{g}}\bm{B}|/\hbar$.
The term beginning with $\Gamma$ consists of any intra-site (IS) spin relaxation (hopping independent).
In a semiclassical picture, the magnetic field rotates the spin orientation $\bm{S}$.
The solution to Eq. (\ref{eq:spinDiffEq}) is
\begin{equation}\label{eq:}
\bm{S}(t) = e^{-\Gamma t}e^{\bm{\Omega}t} \cdot \bm{S}_0 \equiv e^{-\Gamma t}\hat{\bold{R}}(t) \cdot \bm{S}_0,
\end{equation}
where $\bm{S}_0$ is the initial spin vector and $\bm{R}$ is the following rotation matrix:
\begin{equation}\label{eq:rotationMatrix}
\hat{\bold{R}}(t)  = \hat{\bm{1}}+ \sin\omega t \hat{\bm{\Omega}} + 2 \sin^2\frac{\omega t}{2} \hat{\bm{\Omega}}\cdot \hat{\bm{\Omega}}.
\end{equation}
We assume $\bm{S}_0 = S_0\hat{z}$ throughout.
If different spins experience different environments (i.e. different $\bm{B}$), then an average over the different configurations should be taken: $\hat{\bm{R}}_s(t) = \langle \hat{\bm{R}}(t) \rangle$.

This leads us then to address the question of hopping spins.
We approach the problem as a continuous-time random walk.
To introduce the formalism in an intuitive manner, first consider the polarization from an ensemble of stationary spins:
$\bm{P}_0' = e^{-\Gamma t}\hat{\bm{R}}_s(t) \cdot \bm{S}_0$.
Now consider the fact that some of the spins hop to other sites; for the time, let us ignore those spins.
How do we express the stationary spin polarization?
The rotation matrix, $\hat{\bm{R}}_s(t)$, only applies to spins that have not hopped; that fraction is determined by the \emph{survival probability}, $\Phi(t)$.
The survival probability is related to the WTD by the following: 
\begin{equation}
d \Phi(t)/dt = - \psi(t), ~\Phi(t) = \int_t^{\infty} \psi(t')dt' = 1 - \int_0^{t} \psi(t')dt';
\end{equation}
alternatively in Laplace space,
\begin{equation}
\tilde{\Phi}(s) = (1 - \tilde\psi(s))/s.
\end{equation}
The polarization of spins that have not hopped is then $\bm{P}_0 = \hat{\bm{R}}_0(t) \cdot \bm{S}_0$
where a new quantity has been defined as
 \begin{equation}
 \hat{\bold{R}}_{0}(t)\equiv \hat{\bold{R}}_{s}(t)  \Phi(t) e^{-\Gamma t}.
\end{equation}

For a moment let us forget about the stationary spins and examine the behavior of the spins that made the single hop.
For some amount of time these spins were at their home sites and would have experienced  $\hat{\bold{R}}_{s}(t)$.
The amount of rotation depends on the wait-time at the home site; the wait-time is drawn from the WTD.
The amount of rotation in a short time interval $dt'$ is $ \hat{\bold{R}}_{s}(t') \psi(t')dt'$.
At their new site, the spins begin to evolve again with the averaged rotation matrix $\hat{\bold{R}}_{s}(t)$.
If the hop occurred at $t'$ and they precess up to time, $t$, this rotation matrix is described by  $\hat{\bold{R}}_{0}(t-t')$. 
Now we integrate over all possible hopping times to obtain: 
$\int_0^{t} \hat{\bold{R}}_0(t-t')   \hat{\bold{R}}'_{0}(t)dt'$
where we have defined a new quantity:
 \begin{equation}
 \hat{\bold{R}}'_{0}(t)\equiv \hat{\bold{R}}_{s}(t) \psi(t) e^{-\Gamma t},
\end{equation}
which obviously commutes with $\hat{\bold{R}}_{0}(t)$. 

Lastly, we need to include any rotations that might accrue \emph{during} the hop as opposed to before and after the hop.
This matrix, $\hat{\bold{R}}_{h} $, is treated as time-independent and any necessary configurational averaging is assumed.
The total rotation matrix for spins that have hopped once is then
\begin{equation}\label{eq:R1}
\hat{\bold{R}}_{1}(t)= \int_0^{t} \hat{\bold{R}}_0(t-t')   \hat{\bold{R}}_{h} \hat{\bold{R}}'_{0}(t')dt',
\end{equation}
which has the form of a convolution.
Assuming that $\hat{\bold{R}}_{h}$ commutes with the other matrices (most realistically by saying that $\hat{\bold{R}}_{h} \propto \bold{I}$), allows Eq. (\ref{eq:R1}) to be expressed as:
\begin{equation}\label{}
\hat{\bold{R}}_{1}(t)=   \hat{\bold{R}}_{h} \int_0^{t} \hat{\bold{R}}_0(t-t') \hat{\bold{R}}'_{0}(t')dt'.
\end{equation}
The convolution theorem yields
\begin{equation}
\tilde{\hat{\bold{R}}}_{1}(s ) =  \hat{\bold{R}}_{h} \tilde{\hat{\bold{R}}}_{0}(s + \Gamma) \tilde{\hat{\bold{R}}}'_{0}(s + \Gamma).
\end{equation}
The same reasoning is used to find the rotation matrix for spins that have hopped twice:
\begin{equation}
\tilde{\hat{\bold{R}}}_{2}(s ) = \hat{\bold{R}}^2_{h} \tilde{\hat{\bold{R}}}_{0}(s + \Gamma) \tilde{\hat{\bold{R}}}'^2_{0}(s + \Gamma) = \hat{\bold{R}}_{h}\tilde{\hat{\bold{R}}}_{1}(s )  \tilde{\hat{\bold{R}}}'_{0}(s)  .
\end{equation}
The procedure can be continued indefinitely for arbitrary $l$ hops and the following recursive expression is obtained
\begin{equation}
\tilde{\hat{\bold{R}}}_{l}(s ) =  \hat{\bold{R}}^{l}_{h} \tilde{\hat{\bold{R}}}_{0}(s + \Gamma) \tilde{\hat{\bold{R}}}'^{l}_{0}(s + \Gamma)
 =  \hat{\bold{R}}_{h} \tilde{\hat{\bold{R}}}_{l-1}(s) \tilde{\hat{\bold{R}}}'_{0}(s + \Gamma).
\end{equation}
The polarization results from summing this geometric series:
\begin{equation}\label{eq:main}
\tilde{\bold{P}}(s) = \sum_{l=0}^{\infty}\tilde{\hat{\bold{R}}}_{l}(s)\cdot \bold{S}_0 = \tilde{\hat{\bold{R}}}_{0}(s + \Gamma)
[\bm{I} - \hat{\bold{R}}_{h} \tilde{\hat{\bold{R}}}'_{0}(s + \Gamma)]^{-1} \cdot \bold{S}_0.
\end{equation}
It may sometimes also be useful (e.g. if Laplace transform of $\hat{\bold{R}}'_{0}(t)$ has no analytic expression) to write the polarization as a integral equation in the time domain:
\begin{equation}\label{eq:volterra}
\hat{\bold{P}}(t) =  \hat{\bold{R}}_0(t)+ \hat{\bold{R}}_{h}\int_0^{t} \hat{\bold{P}}(t-t') \hat{\bold{R}}'_0(t') dt',
\end{equation}
which is a Volterra equation of the 2nd kind or a renewal equation.

For the sake of pedagogy, we have not yet emphasized the assumptions that lead to our main result, Eq. (\ref{eq:main}).
We now make them clear as they are a subject for discussion in later sections of this article when our results are presented.
We have used a class of assumptions known as the \emph{strong collision approximation}.\cite{Hayano1979}
The approximation has the following characteristics:
(1) local field changes are abrupt and not slow at each hop. 
(2) local fields at or during each hop are uncorrelated from any previous hop.
In the language of stochastic process theory, the evolution is Markovian (2) but not Gaussian-Markovian (1).

A simple example of the theory is demonstrated by using a exponential WTD, $\psi(t) = k e^{-k t}$ where $k$ is the average hopping rate.
Immediately we can write from Eq. (\ref{eq:main}),
\begin{equation}\label{}
\tilde{\bold{P}}(s) =\tilde{\hat{\bold{R}}}_s(s + \Gamma + k)[\bm{I} -k \hat{\bold{R}}_{h} \tilde{\hat{\bold{R}}}_s(s + \Gamma + k)]^{-1} \cdot \bold{S}_0.
\end{equation}
In later sections we show explicit examples of when $\tilde{\bold{P}}(s)$ can be inverted.

\subsection{Multiple Trapping Model}\label{section:MTSection}

More realistic WTDs are much harder to handle. In this article we concentrate on the aforementioned exponential WTD and on the WTD produced by an exponential density of states (which is the appropriate one for amorphous semiconductor band-tails).
The multiple trapping model (MT) constructs a WTD from hopping rates that are of the form of trap release rates, $k(\varepsilon) = k_0 e^{\varepsilon/k_B T}$, where $\varepsilon$ is the trap energy and is distributed exponentially. The energy levels are uncorrelated between hops.
The WTD, $\psi(t)$, is described by $\int_{-\infty}^0 g(\varepsilon)k(\varepsilon)e^{-k(\varepsilon) t} d\varepsilon$ or in Laplace space as
\begin{equation}
\tilde{\psi}(s) = \int_{-\infty}^0 d\varepsilon g(\varepsilon)\frac{k(\varepsilon)}{s + k(\varepsilon)} = 
 \int_{-\infty}^0 dx e^x \frac{k_0 e^{x/\alpha}}{s + k_0 e^{x/\alpha}},
\end{equation}
with $x = \varepsilon/k_B T_0$ and $\alpha = T/T_0$ when using an exponential density of states.
The result can be written as a hypergeometric function:
\begin{equation}\label{eq:hypergeo}
\tilde{\psi}(s) = \frac{k_0 \alpha}{s + \alpha s} ~_2F_1(1,1+\alpha, 2+\alpha, -k_0/s).
\end{equation}
The long time, or asymptotic, behavior of the WTD is of interest. 
It is more straightforward to derive it for the survival probability first;
\begin{equation}
\tilde{\Phi}(s) =
 \int_{-\infty}^0 dx  \frac{e^x}{s + k_0 e^{x/\alpha}} = - \int_{\infty}^0 \frac{dx}{s}  \frac{e^{-x}}{1 + (k_0/s) e^{-x/\alpha}}.
\end{equation}
After making a change of variable, $w = e^{-x} (s/k_0)^{-\alpha}$, we obtain
\begin{equation}
\tilde{\Phi}(s) =
 \int_{0}^{(s/k_0)^{-\alpha}} \frac{dw}{s}\left(\frac{s}{k_0}\right)^{\alpha} \frac{1}{1 + w^{1/\alpha}} .
\end{equation}
Up to now we have not made any assumptions; assuming long times is identical to assuming small $s$ so we can rewrite the WTD as
\begin{equation}
\tilde{\Phi}(s) =
\frac{1}{s} \left(\frac{s}{k_0}\right)^{\alpha} \int_{0}^{\infty} dw \frac{1}{1 + w^{1/\alpha}} .
\end{equation}
The integral is equal to $\pi \alpha \csc(\pi\alpha)$ so the final result in Laplace space is
\begin{equation}
\tilde{\Phi}(s \rightarrow 0) =
\frac{s^{\alpha-1}}{k_0^{\alpha}} \pi \alpha \csc(\pi\alpha) \sim \frac{s^{\alpha-1}}{k_0^{\alpha}},
\end{equation}
where we are not concerned with pre-factors.
The Tauberian theorems dictate that in the time-domain\cite{Klafter2011}
\begin{equation}
\tilde{\Phi}(t \rightarrow \infty) \sim
\frac{t^{-\alpha}}{k_0^{\alpha}}.
\end{equation}
By the identities between the survival probability and the WTD, we can find the asymptotic form of the WTD:
\begin{equation}
\tilde{\psi}(t \rightarrow \infty) \sim
\frac{t^{-\alpha-1}}{k_0^{\alpha}}.
\end{equation}

We stress that the MT ignores correlated hopping that one might suspect in a real system where the energy level the carrier resides is dependent on the previous state. However the WTDs for the two situations actually agree very well.\cite{Hartenstein1996}
The calculated current in the two situations also match which demonstrates that correlations do not contribute heavily to the carrier transport.\cite{Jakobs1993, Hartenstein1996}
We find that correlated hopping can be very important for spin lifetimes and therefore also spin transport.
\begin{figure}[ptbh]
 \begin{centering}
        \includegraphics[scale = 0.255,trim = 45 385 10 80, angle = -0,clip]{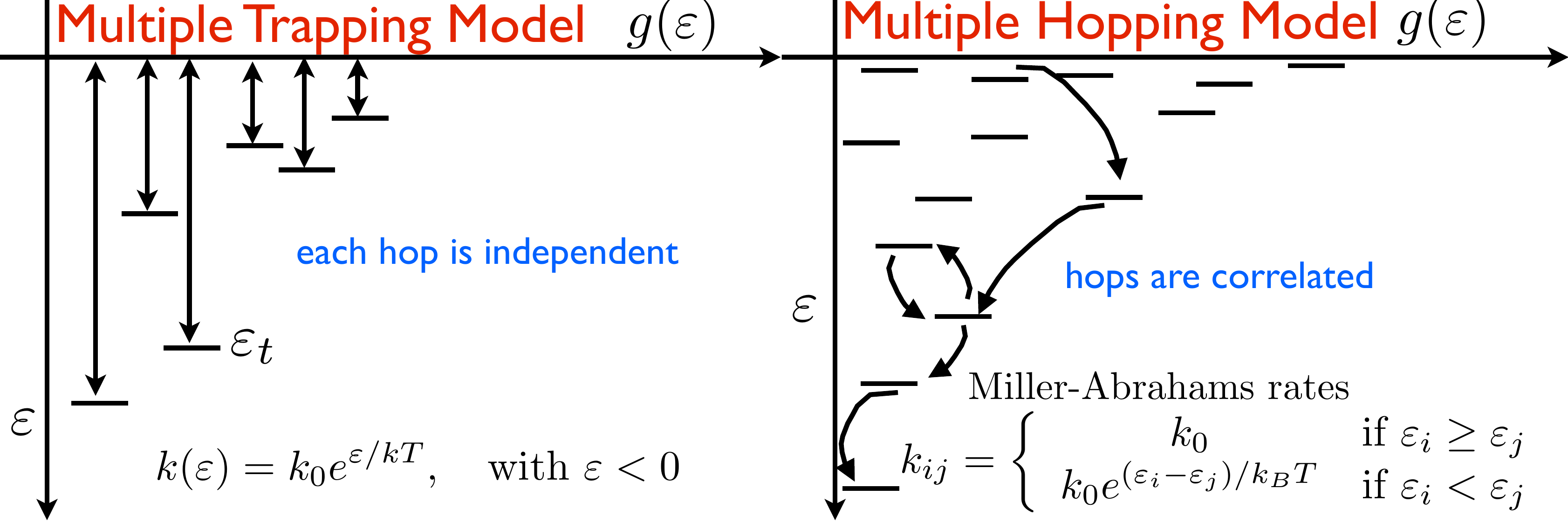}
        \caption[]
{(Color online) Depictions of Multiple Trapping (left) and Hopping (right) models. Each hop in the Multiple Trapping model is independent of its previous hops. We call this uncorrelated hopping. By its very nature then, the local fields felt by the spin are also independent at each hop. The Multiple Hopping models includes correlations. For instance a carrier beginning high in energy (as shown) will tend to cascade downwards in energy when operating under the Miller Abrahams hopping rates. Since sites are correlated, the local fields are also correlated which can be important when a spin hops back and forth between a small number of sites.}\label{fig:MTMH}
        \end{centering}
\end{figure}

\subsection{Multiple Hopping Model}

The Multiple Trapping model treats each hop independently as trap release events with a release rate $k(\varepsilon) =  k_0 e^{\varepsilon/k_B T}$ (see left side of Figure \ref{fig:MTMH}). Such a model ignores the fact that hopping may be correlated to the configuration of sites. For instance, two sites near in energy maybe experience back-and-forth hopping before the charge carrier escapes to some other site.
Therefore treating hops independently as necessitated by Eq. (\ref{eq:main}) is inappropriate in general.
We have chosen to address the more realistic hopping that includes correlations by simulating the spin evolution where hops from $i$ to $j$ are dictated by Miller-Abrahams rates:
\begin{equation}
k_{ij} =  \begin{cases}
  k_0, &  \text{if $\varepsilon_i \geq \varepsilon_j$}  , \\
  k_0 e^{(\varepsilon_i - \varepsilon_j)/k_B T}, &  \text{if $\varepsilon_i < \varepsilon_j$} .
\end{cases}
\end{equation}
Spins are injected randomly into the semiconductor which is modeled as cubic lattice of localizing sites.
The spin of each carrier is sampled at a chosen time interval and averaged over many different configurations of the disorder (typically 10000-50000). A single disorder configuration possesses a fixed landscape of site energies and local fields.

The relevance of correlations (especially for high disorder) for spin transport as opposed to charge transport can be understood by examining site revisitation effects. Carriers oscillate between a small number of sites many times though these sites tend to be near in energy to one another and therefore the oscillations are rapid and do not contribute to the current.\cite{Mendels2013} These oscillations are still spin changing events though and hence can be important for spin relaxation and spin diffusion.\cite{Roundy2013}

Though beyond the scope of this article, correlation effects have been incorporated into CTRW theories on conductivity.\cite{Shlesinger1979}
Applying these methods to the spin diffusion and relaxation problem will be a challenging endeavor for theorists.

\section{The $\delta g$ mechanism for Spin dephasing and Spin decoherence}\label{section:III} 

For simplicity, we consider only isotropic $g$ values such that the $g$-tensor can be written as $\hat{\bold{g}} = g \bold{I}$ where $g$ is a random variable drawn from a Gaussian distribution centered at $g^*$ with width $\Delta g$.
For this case, longitudinal spin relaxation does not exist for this mechanism, so we only examine transverse spin decoherence.
At a given site, $i$, the total angular frequency is expressed as $\bm{\omega}_i = \bm{\omega}_0 + \bm{\delta \omega}_i$ where
\begin{equation}
\bm{\omega}_0 = g^* \frac{\mu_B}{\hbar} B_0\hat{x}, \qquad \bm{\delta \omega}_i = \delta g_i \frac{\mu_B}{\hbar} B_0\hat{x},
\end{equation}
with $B_0$ being the applied field and $\delta g_i$ being the random variable taken from a Gaussian distribution centered at 0 and with standard deviation $\Delta g$.
It is mathematically advantageous to transform to a coordinate system rotating at $-\bm{\omega}_0$ such that the effective angular frequency in the coordinate system at any site is simply $\bm{\delta \omega}_i$.
The spin evolution in the new coordinate system (marked by a prime) is:
\begin{equation}\label{}
\frac{d\bm{S}'(t)}{dt} = \bm{\delta\omega} \times \bm{S}'(t).
\end{equation}
The rotation matrix can be readily found and averaged over the Gaussian distribution of $\delta g$s.
The result is

\begin{eqnarray}\label{eq:transverse}
\bm{R}_s = 
\left( {\begin{array}{ccc}
 1 & 0  & 0  \\
  0 & e^{-a^2 t^2/2}  & 0 \\
 0 & 0  &  e^{-a^2 t^2/2}   \\
 \end{array} } \right), {}
\end{eqnarray}
where $a^2 = (\Delta g ~\mu_B B_0/\hbar)^2$.
In the absence of hopping, the polarization is simply $\bm{P}(t) = \hat{\bm{R}}_s \hat{z}$; the transverse polarization decays in a Gaussian fashion which has been observed in quantum dots and nanocrystals.\cite{Gupta1999, Gupta2002}
In the context of electron spin resonance (ESR) experiments, we label $1/T_2^* = a$.
The line width is Gaussian with width $\Delta B_{1/2} = \Delta g B_0/2$.\cite{Street1991}
It should be remembered that this reduction of polarization is a form of \emph{inhomogeneous dephasing} or broadening and the spin polarization can be recovered by spin echo experiments.

\subsection{The $\delta g$ mechanism for spin decoherence}
\begin{figure}[ptbh]
 \begin{centering}
        \includegraphics[scale = 0.37,trim = 135 175 200 235, angle = -0,clip]{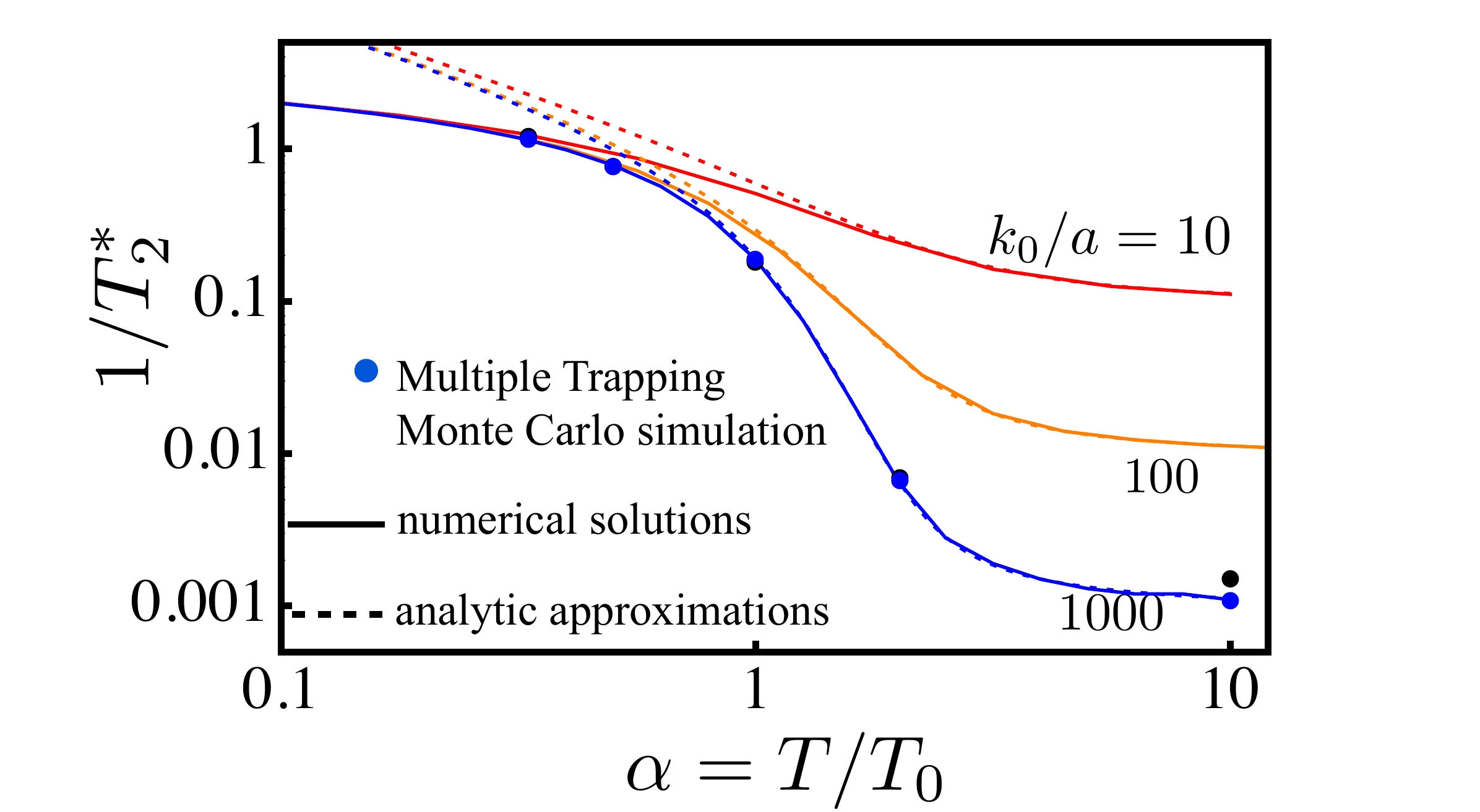}
        \caption[]
{(Color online) Transverse spin relaxation or decoherence rate (in units of $a$) at long times as a function of disorder. Dotted lines are analytic result for $\gamma$, Eq. (\ref{eq:analytic}). Solid lines are numerical solution. Blue solid symbols are fit results from Monte Carlo simulation of Multiple Trapping Model. As $\alpha$ ($a$) increases (decreases), the rate approaches the motional narrowing limit, $a/k_0$, which is expected in the case of low disorder. Black symbols: result from Multiple Hopping Model via Monte Carlo simulations. Multiple Trapping and Hopping models agree well (indistinguishable in plot) except in the low disorder limit of high $\alpha$.}\label{fig:poles}
        \end{centering}
\end{figure}

Once the spins begin to hop the polarization loss is irreversible which is the scenario we examine now.
Using Eq. (\ref{eq:transverse}), reduces Eq. (\ref{eq:main}) to the scalar equation:
\begin{equation}\label{eq:transverse2}
\tilde{P}_z(s) = \frac{\tilde{R}^{zz}_{0}(s )}{
1- \tilde{R}'^{zz}_{0}(s ) }.
\end{equation}
The apparent simplicity of this reduced equation is deceiving. The reason is that we are not dealing with $\tilde{R}'^{zz}_{0}(s) \propto \tilde{\psi}(s)$ but instead $\mathscr{L}[\psi(t) \exp(-a^2 t^2/2)]$ which resists an analytic expression (for WTDs other than the exponential). This Laplace transform is
\begin{eqnarray}\label{}
&&\mathscr{L}[\psi(t) \exp(-a^2 t^2/2)] \nonumber\\
&=& {}\int_{-\infty}^0 \frac{\sqrt{\frac{\pi }{2}} k_0 e^{\frac{\left(k_0 e^{x/\alpha
   }+s\right)^2}{2 a ^2}+\left(\frac{1}{\alpha }+1\right) x}
   \text{erfc}\left(\frac{k_0 e^{x/\alpha }+s}{\sqrt{2} a
   }\right)}{a } dx,
\end{eqnarray}
where $x = \varepsilon/k_B T_0$. 
Using this in Eq. (\ref{eq:transverse2}), the denominator yields only a single pole which dictates that the spin relaxation is exponential at larger times. The pole, which corresponds to the decay rate, can be obtained numerically and is shown as the solid curves in Figure \ref{fig:poles}.
The exact pole can be approximated by expanding the denominator to first order in $s$ and solving for $s$.
The inverse Laplace Transform yields a decay rate
\begin{equation}\label{eq:analytic}
 \frac{I_1 - 1}{I_2},
\end{equation}
where by using $q = e^{x/\alpha}$
\begin{equation}
I_1 = \int_{0}^1 \frac{\alpha  \left(\sqrt{\frac{\pi }{2}} k_0
   q^{\alpha } e^{\frac{k_0^2 q^2}{2 a ^2}}
   \text{erfc}\left(\frac{k_0 q}{\sqrt{2} a
   }\right)\right)}{a }dq
\end{equation} 
and 
\begin{equation}
I_2 = \int_{0}^1 \frac{\alpha  k_0 q^{\alpha } \left(\sqrt{2 \pi } k_0 q
   e^{\frac{k_0^2 q^2}{2 a ^2}} \text{erfc}\left(\frac{k_0
   q}{\sqrt{2} a }\right)-2 a \right)}{2 a ^3} dq
\end{equation} 
which can be written in close form in terms of generalized hypergeometric functions.
As seen in Figure \ref{fig:poles} (dotted curves), the adequacy of the approximation hinges on the value of $\alpha$; as $\alpha$ gets smaller than unity, the approximation is completely inadequate.
In the limit of fast hopping/low disorder, the decoherence rate, Eq. (\ref{eq:analytic}), approaches the motional narrowing value of $1/T_2^* = a^2/k_0$.
Lastly Figure \ref{fig:poles} also shows the results from fitting the Monte Carlo simulation of the Multiple Trapping problem (blue solid symbols). Agreement in relaxation rates between the simulation (Multiple Hopping Model) and theory (Multiple Trapping Model) is very good except in the regime of low disorder where site revisitation effects are important. The next section discusses this topic in greater detail.

\begin{figure}[ptbh]
 \begin{centering}
        \includegraphics[scale = 0.37,trim = 135 165 200 235, angle = -0,clip]{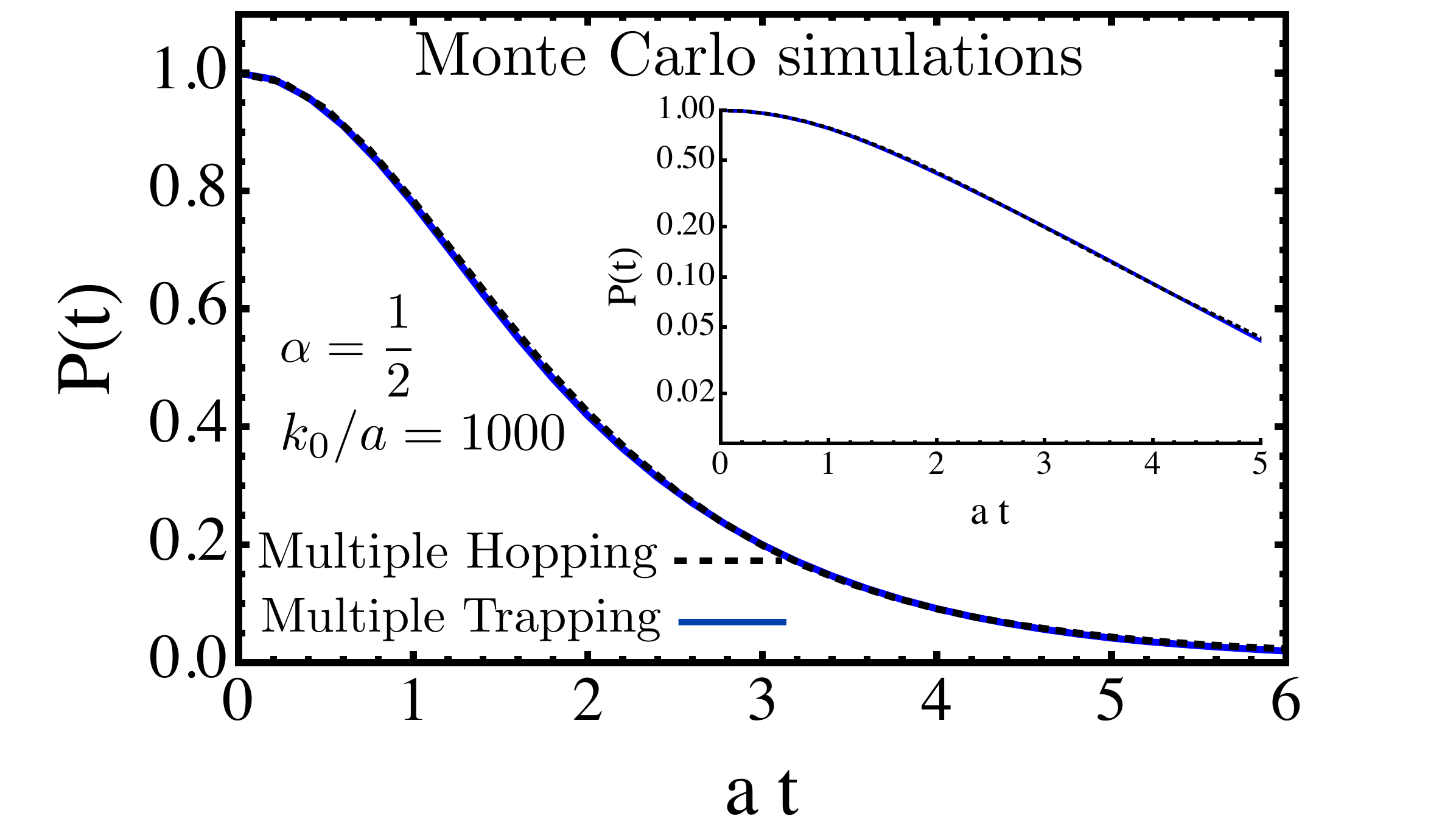}
        \caption[]
{(Color online) Spin polarization function versus time as determined by Monte Carlo simulations. In the disordered case ($\alpha < 1$), the Multiple Trapping model agrees well with the more realistic Multiple Hopping model. Short times possess a Gaussian type decay while longer times exhibit exponential relaxation. Inset: plot on logarithmic scale showcasing exponential decay at longer times.}\label{fig:hiD}
        \end{centering}
\end{figure}
An example of the polarization function's shape is shown in Figure \ref{fig:hiD}. 
For the slowest hopping/largest disorder, the polarization decays in a Gaussian fashion as Eq. (\ref{eq:transverse}) makes clear.
As the hopping rate is continually increased or the disorder decreased, the polarization function develops more exponential character until the motional narrowing regime is obtained.

\subsection{The role of correlations}

In light of Figure \ref{fig:poles}, we see that the difference between the Multiple Hopping and Trapping calculations are minimal at intermediate to large disorder strengths ($\alpha \lesssim 1$) for the hopping $k_0 = 1000 a$ (the same is true for smaller $k_0$ also).
In this regime correlations between site energies and local fields must be inconsequential.
The reason for this is the following: 
for large disorder, wait-times are typically longer than the local field period such that $\tau_h a \gg 1$. The transverse spin ensemble then decoheres on a time scale of the order of $1/a$ which is what we observe to be happening in Figure \ref{fig:poles}. 
Whether a spin is frequenting a site often is irrelevant since the phase of the spin is randomized by the time it embarks on its first hop.
The same reasoning can be used to explain the equivalence of Multiple Hopping and Trapping models when calculating hyperfine spin relaxation.

The discrepancy between the Multiple Hopping and Trapping Models is largest when disorder is small and hopping is rapid (\emph{e.g.} at $\alpha = 10$ in Figure \ref{fig:poles}).
Since spin decoherence times are much longer in this regime, the role of correlations is larger and apparent.
Rotations in one dimension (only considering transverse decoherence here) commute.
So returning to a particular site one time (each stay being of length $\tau_i$ and $\tau_j$) is equivalent to having stayed at the site for a duration $\tau_i + \tau_{j}$ and not returned to it.
From the theory of random walks, the mean number of visits to each site of a simple cubic lattice is 1.516.\cite{Montroll1965, Czech1989b}
The mean wait-time is increased by this amount $1/k_0 \rightarrow 1.516/k_0$.
\cite{Czech1989}
The motional narrowing spin relaxation rate is modified to be $1/T_2^* = 1.516 a^2/k_0$.
Our Multiple Hopping simulations recover this result in the low disorder limit as shown at $\alpha = 10$ in Figure \ref{fig:poles} (black symbol).

\section{Spin-orbit Spin Relaxation}\label{section:IV} 

Each hop brings about a sudden spin-rotation given by the rotation matrix of Eq. (\ref{eq:rotationMatrix}) except that now the rotation angles are independent of time; this fact simplifies the mathematics considerably. 
 For simplicity we make the following assumption:  the spin-orbit field components are distributed as a Gaussian function with width $\gamma$.
 For the MT model, we are interested in the spatially averaged $\hat{\bm{R}}_h$ which, given the assumed isotropy of the spin-orbit fields, is proportional to the identity matrix. 
In the small angle approximation, the averaged rotation matrix is simply
\begin{equation}
\hat{\bm{R}}_h = (1 - \gamma^2)\hat{\bm{1}}.
\end{equation}
Equation \ref{eq:main} then reduces to a more manageable form:
\begin{equation}\label{}
\tilde{P}_z(s) = \frac{\tilde{\Phi}(s )}{
1- (1 - \gamma^2)\tilde{\psi}(s ) }.
\end{equation}
For the special case of the exponential WTD, the Laplace inversion is determined exactly to yield $P_z(t) = e^{- \gamma^2 k t}$ which is in agreement with existing theories of spin-orbit spin relaxation.\cite{Elliott1954, Yu2011} 
Within the hitherto described Multiple Trapping model, we can express the polarization in terms of special functions by using Equation (\ref{eq:hypergeo}).
The polarization in time can be ascertained by numerically inverting the Laplace transform.\cite{Valko2004, Abate2004} 
However in the long-time case, the asymptotic analysis of Section \ref{section:MTSection} can be used to find an analytic expression:
\begin{equation}
P(t) = \frac{1}{\Gamma(1-\alpha)} \frac{1}{k_0^{\alpha}}\frac{1}{\gamma^2} \pi \alpha \csc(\pi \alpha) t^{-\alpha}
\end{equation}
which shows that spin polarization is characterized by \emph{algebraic} decay.\cite{Harmon2013a}
Some physical intuition regarding this result can be obtained by noting that $P(t) = \Phi(t)/\gamma^2$.
By recalling that $\Phi(t)$ is the probability that a hop has \emph{not} taken place up to time $t$, we see that the polarization decays as carriers hop in agreement with what is true from low disorder case.
Thus the transport of spin is inherently detrimental to spin preservation as expected from the similar Elliott-Yafet mechanism in inorganic semiconductors.\cite{Elliott1954, Yafet1963}
\begin{figure}[ptbh]
 \begin{centering}
        \includegraphics[scale = 0.325,trim = 85 100 200 180, angle = -0,clip]{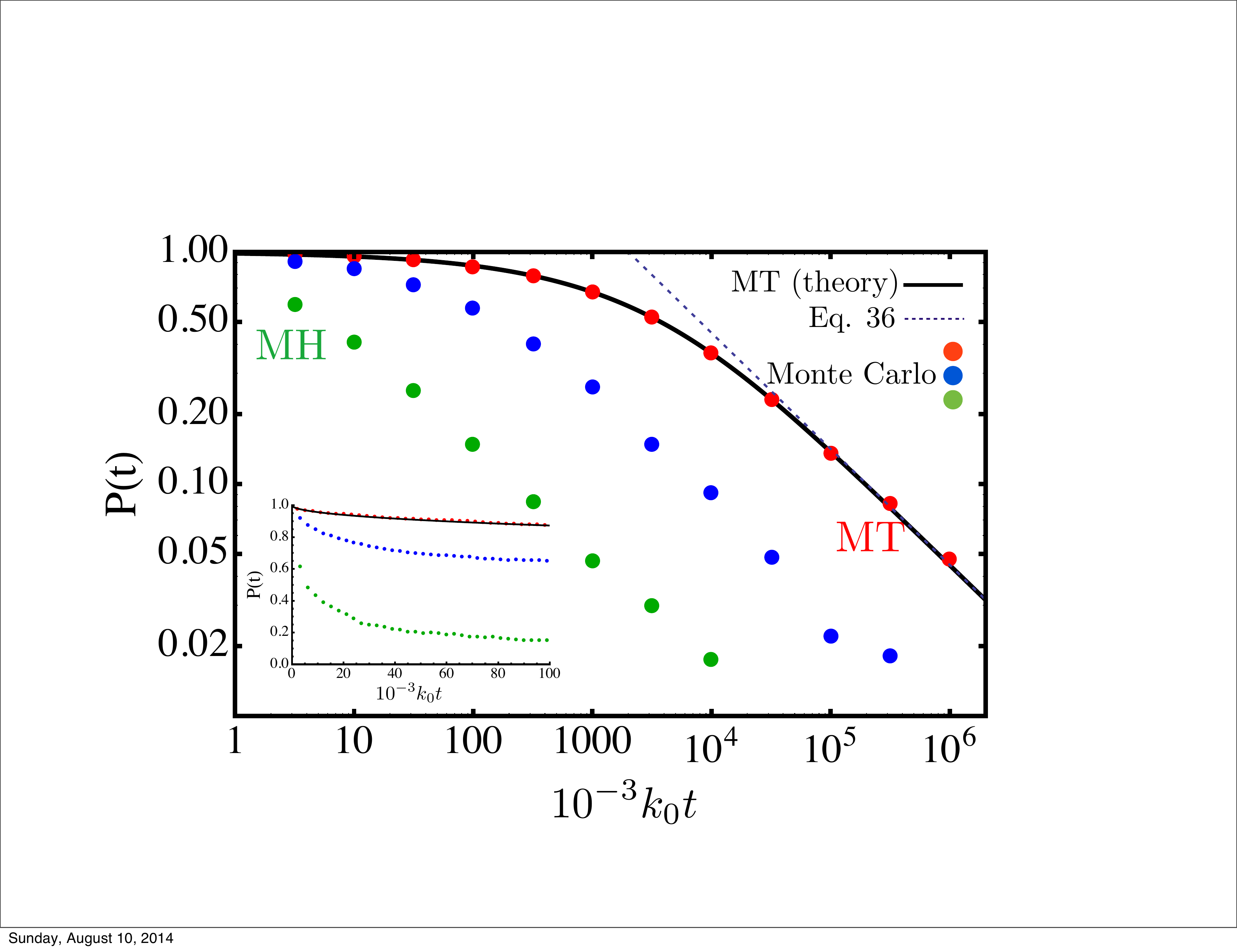}
        \caption[]
{(Color online) The spin polarization as a function of time when relaxation is due to spin-orbit coupling using $\alpha = 1/2$, $\gamma = 0.025$. Inset: same data but axes are on a linear scale and the time scale is shorter. Carriers are injected at sites randomly in the semiconductor.}\label{fig:SOPlot}
        \end{centering}
\end{figure}

Figure \ref{fig:SOPlot} displays the results of our CTRW theory (black solid line from numerical Laplace inversion) and analytic asymptotic expression (dotted line).
A large discrepancy appears between the Multiple Trapping and Hopping models, though the qualitative features (algebraic decay) are identical.
Unlike what was found for the $\delta g$ mechanism, correlations are much more pivotal to the SOC mechanism.

The three curves of Figure \ref{fig:SOPlot} depict the full Multiple Hopping result (green symbols), the Multiple Trapping result (red symbols and solid black line), and the result where correlated hopping exists but local fields are uncorrelated (blue symbols). The dramatic differences between the long-time scales of the three curves indicates the importance of correlated hopping and correlated fluctuating fields though the algebraic dependence is retained in all three.

The indicated results assume spins are injected randomly into the amorphous semiconductor (i.e. no site-energy dependence). 
We find that the spin lifetime is contingent on the injection conditions.
For example if the spins are injected preferably to sites lower in energy, the time to decay is significantly lengthened  since the the rapid cascading at early times is avoided (not shown).
This observation suggests that tuning the spin injection (by a bias perhaps) could alter the spin relaxation times.

\section{Spin relaxation with localized and extended states}\label{section:V} 

Up to this point, we have only concerned ourselves with spin relaxation for spin carriers hopping within the mobility gap.
However one should also account for scenarios that involve carriers hopping up to the more conductive states above the mobility edge.
The CTRW and multiple trapping theories, already previously introduced here, can be straightforwardly extended to this more complicated situation.

In the following section, we treat the problem generally where the two subsystems (above and below mobility edge) are not yet specified. 

\subsection{Spin Relaxation of Carriers that Cross between Two Systems}

Consider two subsystems, $E_0$ and $E_1$, that possess their individual set of spin interactions that will lead to spin relaxation and decoherence which can be denoted by polarization matrix functions, $\hat{\bm{P}}_0(t)$ and $\hat{\bm{P}}_1(t)$ (these would be the type of functions calculated in the previous sections), and also their particular WTDs, $\psi_0(t)$ and $\psi_1(t)$. However, the two subsystems are not closed from one another; there is intersystem crossing that is not necessarily symmetric. Environment-specific WTDs can be defined: $\psi_{0\rightarrow1}$ and $\psi_{1\rightarrow0}$ which give  the distribution of wait-times before the carrier transitions from $E_0$ to $E_1$ and vice-versa, respectively. $\Phi_{0}(t)$ and $\Phi_{1}(t)$ are survival probabilities for remaining in $E_0$ and $E_1$, respectively. A final ingredient is possible spin rotations incurred while crossing systems; traveling from $E_0$ to $E_1$ gives $\hat{\bm{R}}_{0\rightarrow1}$ and the opposite holds for the reverse transition. For simplicity, we assume these rotation matrices to be isotropic (diagonal) and independent of time.

Consider the polarization function of a particle that is initiated in $E_1$ and at some later time, $t$, is also found in $E_1$ though any number of intersystem crossings can occur between $0$ and $t$.
We call such a function $\hat{\bm{Q}}_{ 1\rightarrow 1}$ and determine it in the following way:
\begin{widetext}
\begin{eqnarray}
\hat{\bm{Q}}_{1 \rightarrow 1} (t)= 
\Phi_{1}(t) \hat{\bm{P}}_1(t) +  \hat{\bm{R}}_{0\rightarrow1}  \hat{\bm{R}}_{1\rightarrow0}\int_0^t dt' \int_0^{t'}dt'' \Phi_{1}(t-t') \hat{\bm{P}}_1(t-t') \psi_{0 \rightarrow 1}(t'-t'') \hat{\bm{P}}_0(t'-t'') \psi_{1 \rightarrow 0}(t'')\hat{\bm{P}}_1(t'') + ...
\end{eqnarray}
\end{widetext}
where each additional term introduces two more intersystem transitions.
As before, it is advantageous to transform to the Laplace domain:
\begin{widetext}
\begin{eqnarray}
\tilde{\hat{\bm{Q}}}_{1 \rightarrow 1} (s) &=&
\tilde{\hat{\bm{V}}}_1(s) +   \tilde{\hat{\bm{V}}}_1(s)  \tilde{\hat{\bm{V}}}'_0(s)  \tilde{\hat{\bm{V}}}'_1(s) +   \tilde{\hat{\bm{V}}}_1(s)  \tilde{\hat{\bm{V}}}'_0(s)  \tilde{\hat{\bm{V}}}'_1(s) \tilde{\hat{\bm{V}}}'_0(s)  \tilde{\hat{\bm{V}}}'_1(s) + ...\nonumber\\
& =& {} \tilde{\hat{\bm{V}}}_1(s)  \sum_{n = 0}^{\infty}\big[  \tilde{\hat{\bm{V}}}'_0(s)  \tilde{\hat{\bm{V}}}'_1(s) \big]^n = \tilde{\hat{\bm{V}}}_1(s)  \big[ \hat{\bm{1}} -   \tilde{\hat{\bm{V}}}'_0(s)  \tilde{\hat{\bm{V}}}'_1(s)\big]^{-1}
\end{eqnarray}
\end{widetext}
where $\tilde{\hat{\bm{V}}}_i(s) = \mathscr{L}[\tilde{\Phi}_{i}(t)\hat{\bm{P}}_i(t) ] $ and $\tilde{\hat{\bm{V}}}'_i(s) = \hat{\bm{R}}_{i \rightarrow \neq i} \mathscr{L}[\tilde{\psi}_{i \rightarrow \neq i}(t)\hat{\bm{P}}_i(t) ] $.
By symmetry, another contribution can be readily expressed as
\begin{equation}
\tilde{\hat{\bm{Q}}}_{0 \rightarrow 0} (s) = \tilde{\hat{\bm{V}}}_0(s)  \big[ \hat{\bm{1}} - \tilde{\hat{\bm{V}}}'_1(s)  \tilde{\hat{\bm{V}}}'_0(s)\big]^{-1}.
\end{equation}
The other two contributions are found by modifying $\tilde{\hat{\bm{Q}}}_{0 \rightarrow 0} (s) $ and $\tilde{\hat{\bm{Q}}}_{1 \rightarrow 1} (s) $.
For $\tilde{\hat{\bm{Q}}}_{0 \rightarrow 1} (s) $, the first term in the series involves one transition and the final state is not identical to the initial state so
\begin{equation}
\tilde{\hat{\bm{Q}}}_{0 \rightarrow 1} (s) = \tilde{\hat{\bm{V}}}_1(s) \tilde{\hat{\bm{V}}}'_0(s)  \big[ \hat{\bm{1}} - \tilde{\hat{\bm{V}}}'_1(s)  \tilde{\hat{\bm{V}}}'_0(s)\big]^{-1}.
\end{equation}
Likewise,
\begin{equation}
\tilde{\hat{\bm{Q}}}_{1 \rightarrow 0} (s) = \tilde{\hat{\bm{V}}}_0(s) \tilde{\hat{\bm{V}}}'_1(s)  \big[ \hat{\bm{1}} - \tilde{\hat{\bm{V}}}'_0(s)  \tilde{\hat{\bm{V}}}'_1(s)\big]^{-1}.
\end{equation}
If $c$ fraction of spins start in $E_0$ then the total polarization matrix in the time domain is
\begin{equation}
\hat{\bold{P}}(t) = (1-c)\big[\hat{\bold{Q}}_{1\rightarrow1}(t) +\hat{\bold{Q}}_{1\rightarrow0}(t) \big] + c\big[\hat{\bold{Q}}_{0\rightarrow0}(t) +\hat{\bold{Q}}_{0\rightarrow1}(t) \big],
\end{equation}
which can be determined by numerical Laplace inversion if the $\hat{\bold{Q}}_{i\rightarrow j}$ can be expressed in Laplace space.

\subsection{Spin-Orbit Spin Relaxation from Intercrossing between Extended States and Localized States at a Single Energy Level}

The simplest example of the intercrossing is the case where the Fermi energy, $\varepsilon_F$ (in units of $k_B T$ and defined with respect to $\varepsilon_c = 0$), lies in the band tail. 
To avoid the complexity of different energy states in the band tail, we consider only those spins at the Fermi level. To be released from the localized state and into the highly conductive states, an energy $-\varepsilon_F$ is required. The rate for this to happen is then $k_r e^{\varepsilon_F}$. Conversely, a itinerant spin is occasionally trapped back down to the localized state at a rate $k_t$ (we assume exponential WTDs for both the trapping and release processes).
For both environments, we examine only the transport-induced spin relaxation which is primarily from the SOC. 
Thus, 
\begin{eqnarray}
&&P_0(t) = e^{-\gamma_0^2 k_0 t}, ~P_1(t) = e^{-\gamma_1^2 k_1 t}, ~ c = 0 {}\nonumber\\
&&\psi_{0\rightarrow 1}(t) = k_r e^{\varepsilon_F} e^{-k_r e^{\varepsilon_F} t},~ \psi_{1\rightarrow 0}(t) = k_t e^{-k_t t}{}\nonumber\\
&&\hat{\bm{R}}_{0\rightarrow 1} = (1-\gamma_{01}^2)\hat{\bm{1}}, ~\hat{\bm{R}}_{1\rightarrow 0}  =(1-\gamma_{10}^2)\hat{\bm{1}}{}
\end{eqnarray}

From these definitions, calculating the $\tilde{\hat{\bm{Q}}}_{i\rightarrow j}$(s) is straightforward.
The total polarization can be inverted though the final expression is quite cumbersome. Most important for our purposes are the two spin relaxation rates that can be extracted from the exponential decay:
\begin{widetext}
\begin{equation}
\Gamma_{fast/slow} = 
\frac{1}{2} \left(k_r e^{\varepsilon_F}  + \gamma^2  (k_0 + k_1)   + k_t \pm \sqrt{\left( k_r e^{\varepsilon_F} + \gamma^2  (k_0 + k_1)+  k_t\right)^2-4 \gamma^2
   \left(k_0(k_t + \gamma^2 k_1) + k_r e^{\varepsilon_F} (k_1 + k_t (2-\gamma^2))\right)}\right)
\end{equation}
\end{widetext}
where we stipulated that all spin-orbit parameters are equal, $\gamma_0 = \gamma_1 = \gamma_{01} = \gamma_{10}$.
The spin relaxation among the localized states should be much smaller than in the extended states. Also trap release times could be significantly longer than trapping times. 
Expanding the rates with these reasonable assumptions in mind yields
\begin{equation}\label{eq:slow}
\Gamma_{fast/slow} =  \begin{cases}
  \gamma^2 k_1 + k_t &  \text{}  \\
  \frac{\gamma^2 k_1+ 2 \gamma^2  k_t}{\gamma^2 k_1 +   k_t} k_r e^{\varepsilon_F} &  \text{} .
\end{cases}
\end{equation}
If the itinerant spin relaxation rate is sufficiently large (i.e. $\gg k_t$), then the slow rate simplifies to $k_r e^{\varepsilon_F}$ which indicates that localized spins that are promoted to the higher conductive states typically lose their orientation before they can be trapped again.
Despite the fact that the fast rate remains in the absence of SOC, we have checked the full solution to be $(\hat{\bm{Q}}_{11}(t) + \hat{\bm{Q}}_{10}(t))\cdot \hat{z} = 1$ if $\gamma = 0$ as expected.

\section{Amorphous Inorganic Semiconductors}\label{section:VI}

The theory just outlined is now applied to the case of spin polarized carriers in amorphous semiconductors. In this section the basic properties of these materials are summarized. In the following subsections, we compare the result of the theory to available ESR experiments.

\begin{figure}[ptbh]
 \begin{centering}
        \includegraphics[scale = 0.35,trim = 145 270 190 60, angle = -0,clip]{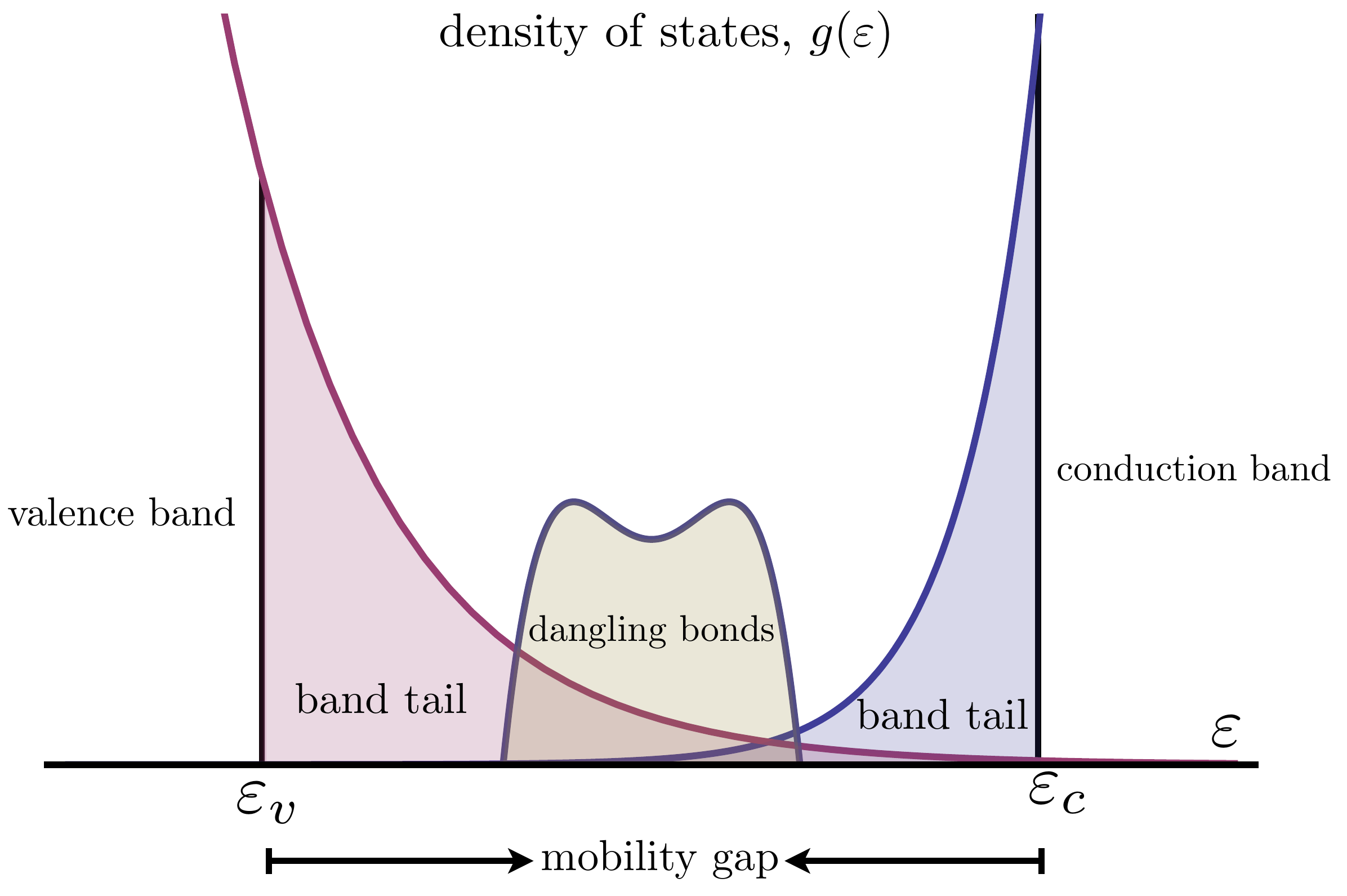}
        \caption[]
{Density of states for a-Si.}\label{fig:Fig1}
        \end{centering}
\end{figure}

Transport properties of amorphous semiconductors are characterized by two mobility edges at energies, $\varepsilon_v$ and $\varepsilon_c$, outside of which states are extended (but still unlike Bloch waves) and between which states are localized. These localized states are said to fall within the ``mobility gap" of the amorphous semiconductor. Refer to Figure \ref{fig:Fig1}.
The localized states are intrinsic to the semiconductor and not necessarily the result of impurities.
There are two types of defect states in pure a-Si: dangling bonds which  lie near the center of the band gap and band tail states that are formed from the distorted nature of the lattice by way of variations in bond lengths, bond angles, and dihedral angles.
The former energy states are believed to vary not nearly as rapidly as the latter tail states which possess, near the mobility edges, a density of these states exponential in nature.\cite{Fuhs2006} For example, below $\varepsilon_c$ these so-called ``band-tail" states are given by
\begin{equation}
g(\varepsilon) = \frac{1}{k_B T_0} e^{\varepsilon/k_B T_0},
\end{equation}
where 
$k_B$ is Boltzmann's constant, $T_0$ is the distribution's width in Kelvin and a measure of the disorder, and energy $\varepsilon$ is taken to be negative. 
An analogous expression exists for the valence band tail density of states.
Typical values for $T_0$ are in the range of 300-600 K; $T_0$ varies between the valence and conduction mobility edges and tends to be larger for the valence band tail.
We assume the dangling bond density of states is approximately constant. \cite{Movaghar1977, Dersch1981b}

Commonly, the Fermi energy lies within the mobility gap; we assume electron majority carriers though the theory applies equally well to hole majority carriers. 
Pure a-Si possesses a large dangling bond density of states which effectively pins the Fermi level near mid band gap which ultimately makes the electronic properties of these materials immune to doping. By passivating the dangling bonds through the incorporation of hydrogen, the density of mid gap states is reduced and the hydrogenated material, a-Si:H, can be successfully either $p$ or $n$-doped. 
Amorphous silicon is prepared through a variety of methods, the most prevalent being plasma-enhanced chemical vapor deposition (PECVD). 
In this process, silane (SiH$_4$) can be added whose hydrogens eventually passivate a portion of the silicon dangling bonds.
Additionally PECVD allows for the incorporation of phosphine (PH$_3$) and diborane (B$_2$H$_6$) which lead to $n$-doping from phosphorous and $p$-doping form boron.
The superior properties of a-Si:H led to the material being utilized in electronic devices.

\subsection{Spin properties}

Amorphous semiconductors such as a-Si and a-Ge also contain few nuclear spins so we can reasonably expect SOC to be limiting. Their hydrogenated counterparts (a-Si:H and a-Ge:H) do obviously have significantly more nuclear moments though the observed affect on the ESR line width is surprisingly minimal which indicates the paramagnetic dangling bonds are well isolated from the hydrogen and are highly localized on the silicon atoms.\cite{Taylor1984}
For these reasons we will not further explore hyperfine induced spin relaxation in this article.
IS mechanisms, that are independent of mobility, have been studied in some detail in the past\cite{Stutzmann1983} but will not be delved into here as our focus is on transport-induced spin relaxation which is often observed to be the dominant source of spin relaxation at room temperature.

Additionally, the heavier elements present (Si and Ge) suggest SOC effects to be greater than that found in organic materials.
The SOC strength of $\gamma\approx 0.1$ has been used often in the literature.\cite{}
This value is about threes time larger than that found for the oft-studied Alq$_3$ organic semiconductor.\cite{Yu2011, Yu2012}
The larger SOC also gives rise to inhomogeneous g-factors which can dephase or decohere spins in an applied field.
This effect - which has been considered negligible in organic semiconductors - has been observed to contribute significantly to ESR line widths in amorphous semiconductors below room temperature.\cite{Movaghar1978, Biegelsen1980}

In the following sections below we discuss spin relaxation in three regimes of inorganic semiconductors (specifically a-Si or a-Si:H):
1) hopping within dangling bond states
2) hopping within band tail states
3) trapping and activation above and below the mobility edges at $\varepsilon_v$ or $\varepsilon_c$.
Low occupational probabilities and low spin injection densities allow us to assume dilute carriers and avoid complicating features such as dipolar and exchange interactions.\cite{Bachus1979, Biegelsen1980, Taylor1984, Yu2013a}

\subsection{Undoped a-Si}

In undoped a-Si, the Fermi energy lies near mid band gap within a large density of states coming from dangling bonds. Charge transport in this regime occurs via variable range hopping.\cite{Thomas1981}
Using the results of Section \ref{section:III} and the observed line width, $\Delta B_{1/2} = \Delta g B_0/2 = 7.5$ G, where $B_0$ is the field corresponding to a resonant frequency of 9 GHz, a $\Delta g \approx 5 \times 10^{-3}$ is ascertained.\cite{Lee1973, Thomas1978, Street1991}
\begin{figure}[ptbh]
 \begin{centering}
        \includegraphics[scale = 0.4,trim = 180 252 312 242, angle = -0,clip]{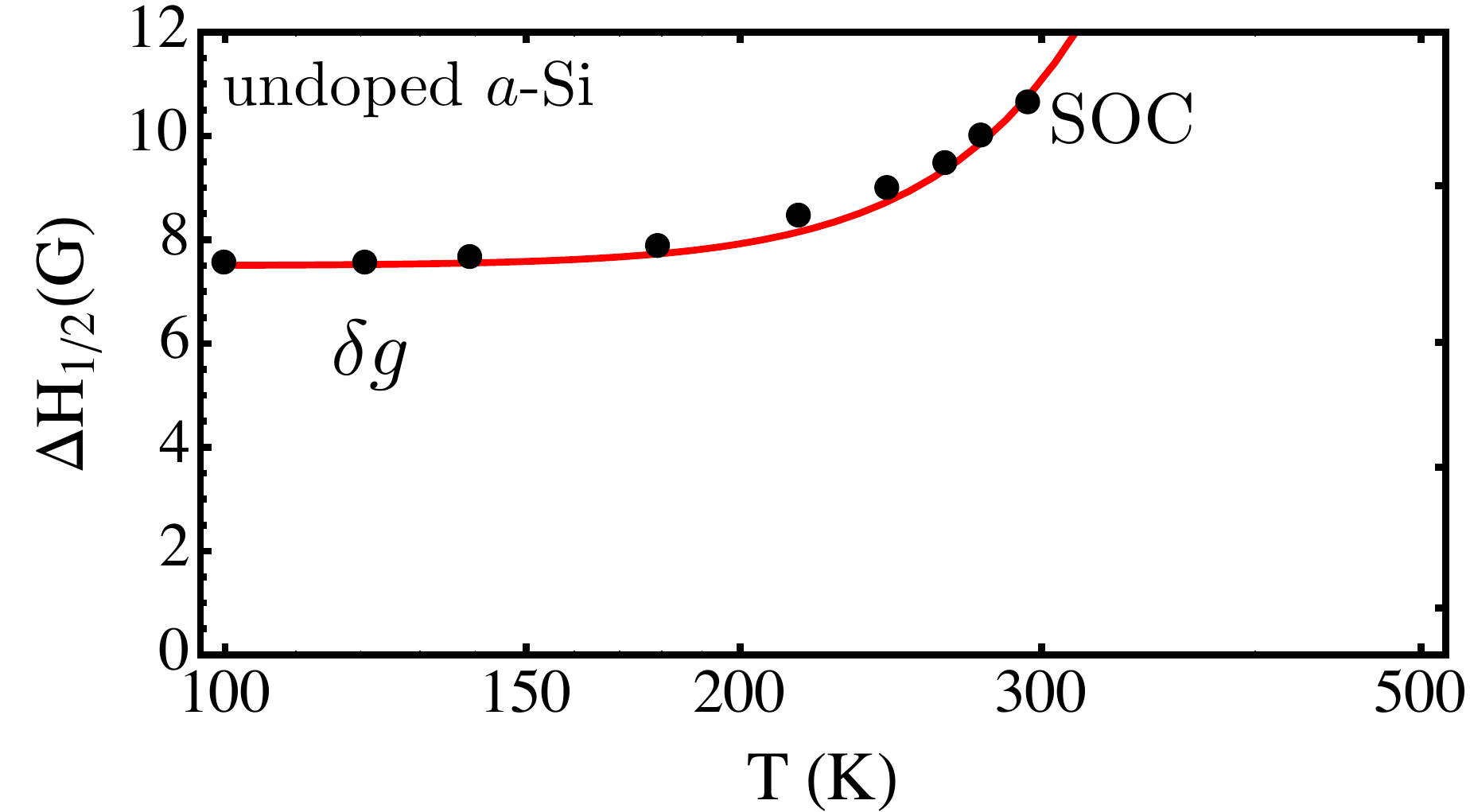}
        \caption[]
{(Color online) Black solid symbols are experiments of Refs. \onlinecite{VogetGrote1976} and \onlinecite{Movaghar1977}. Red solid line is our theoretical result. Parameters used: $\gamma = 0.1$, $k_0 = 10^{10}$ ns$^{-1}$, $T_0 = 4.8\times 10^{7}$ K. Other experimentalists have seen the same variable range hopping dependence for line width.\cite{Bachus1979}}\label{fig:movaghar}
        \end{centering}
\end{figure}
The temperature dependent portion is found from our non-dispersive result in Section \ref{section:IV}, $1/T_2 = \gamma^2 k(T)$ which is a result first realized by Movaghar and Schweitzer in Ref. \onlinecite{Movaghar1977}. The dangling bond density of states is slowly varying around the Fermi level so $k(T) = k_0 \exp(-(T/T_0)^{1/4})$ under the assumption of variable range hopping within a constant density of states. Relaxation times in this regime are on the order of one nanosecond at room temperature.

\subsection{p-doped a-Si:H (trapping and releasing around the mobility edge)}

\begin{figure}[ptbh]
 \begin{centering}
        \includegraphics[scale = 0.4,trim = 225 95 200 120, angle = -0,clip]{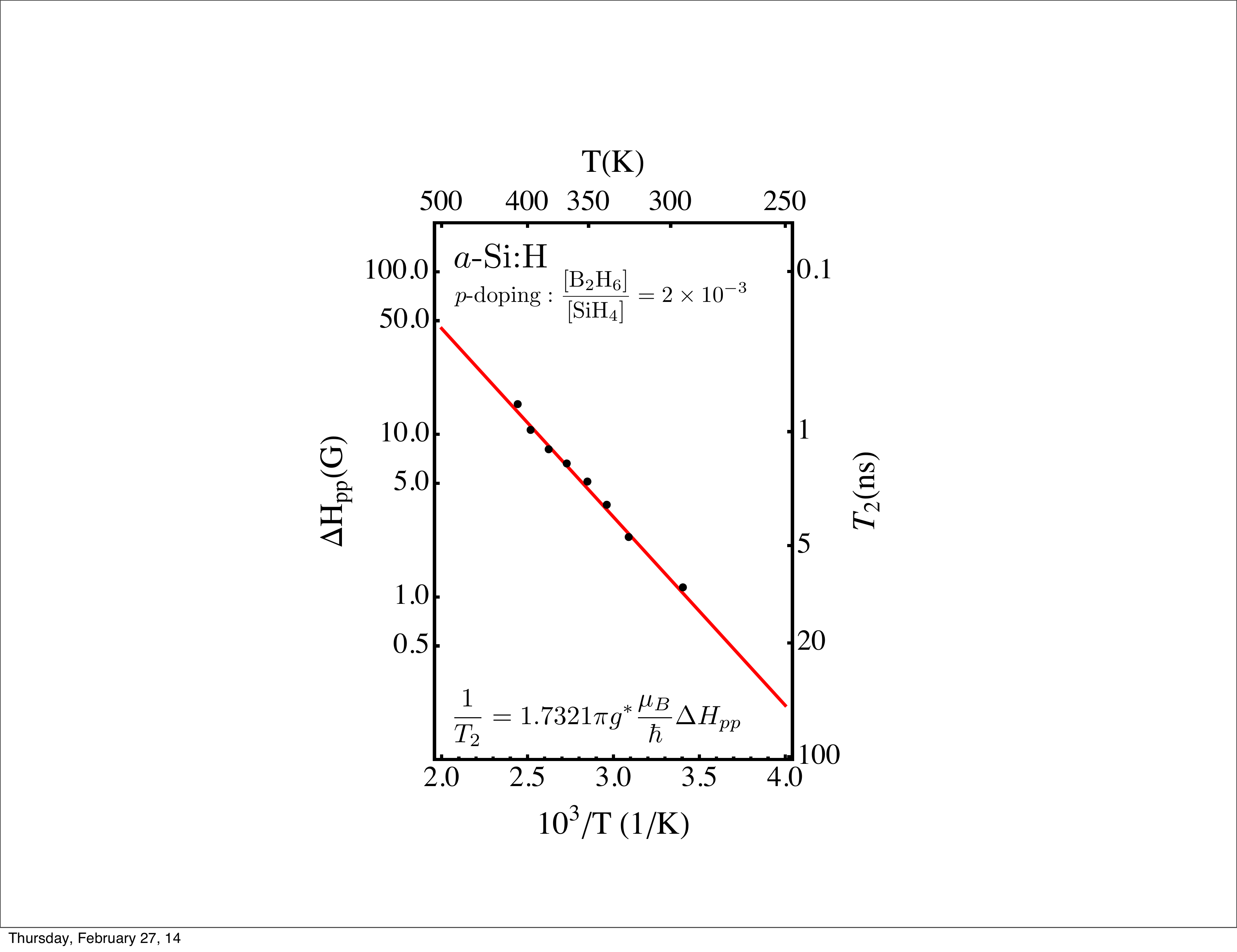}
        \caption[]
{(Color online) Black solid symbols are experiments of Refs. \onlinecite{Dersch1981a} and \onlinecite{Dersch1981b} as reported in Ref. \onlinecite{Overhof1982}. Red solid line is our theoretical result using the slow relaxation rate of Eq. (\ref{eq:slow}) using $k_r = 9 \times 10^2$ ns$^{-1}$ and $\Delta E = -0.23$ eV.}\label{fig:Overhoff}
        \end{centering}
\end{figure}

Using the slow rate of Eq. (\ref{eq:slow}) yields a rate $1/T_2 \approx k_r \exp(\Delta E)$ if $k_1 \gg k_t$ which is shown in Figure \ref{fig:Overhoff} along with experimental data on $p$-doped a-Si:H.
We have used a generic activation $\Delta E$ instead of $\varepsilon_F$ because in actuality the mobility edge is ambiguously defined due to the existence of long-ranged electrostatic potentials from negatively charged acceptors which effectively shift the Fermi level by some amount; knowledge of the Fermi level is also obscured by its temperature dependence.\cite{Overhof1981, Overhof1982, Fuhs2006}
These ambiguities aside, independence of the spin relaxation to the SOC and the fast hopping rate $k_1$ indicates that the spin relaxation is controlled by the release of carriers from the slowly relaxing localized states to the fast relaxing itinerant states.

Another mechanism that could play a role is fast spin exchange between localized and extended states even if the extended states are sparsely populated.\cite{Pines1957, Feher1959, Dersch1981b, Overhof1982}
We do not investigate this process here.

\subsection{doped a-Si:H (transport in band tail)}

Lastly, we consider doped a-Si:H such that the Fermi level lies in either the conduction or valence band tail (i.e. exponential density of states).
The theory described in previous sections dealing with an exponential density of states assumed a dilute limit of carriers which is the scenario present in time-of-flight  (and spin injection)  experiments where dispersive transport is observed.
This assumption was also implicit in our simulations of a carrier hopping among completely empty states.
A feature of this system, and one that distinguishes it from faster decaying density of states, is that the average energy of the carrier continually dives in energy.\cite{Baranovskii2014}
The algebraic spin relaxation predicted herein may be difficult to measure for the following reasons: the slow relaxation will be masked by other faster mechanisms; in real systems spin-spin interactions between charges must be accounted - exchange between quasi-stationary spins and faster hopping spins will tend to reduce the breadth of relaxation times that give rise to the algebraic decay.\cite{Overhof1981}

In real systems, carriers dive until they eventually reach energies near the Fermi level in which case occupations effects now play a role.\cite{Oelerich2012}
In these equilibrium situations, the WTDs used in our theory must be defined with care taken to the occupation of states.\cite{Bobbert2009}
Such calculations are beyond the scope of this article.

\section{Conclusions}

We have presented a continuous-time random walk theory of spin relaxation that is applicable in variety of systems that display incoherent charge transport.
The theory can account for any number of relaxation mechanisms though we have chosen to focus particularly on relaxation emanating from spin-orbit effects in inorganic semiconductors. When applying the model to amorphous semiconductors, we find excellent agreement with ESR data. Our random walk theory also predicts new spin relaxation regimes (algebraic spin decay) for charge transport within amorphous semiconductor band tails.

\section{Acknowledgements}
This work was supported in part by C-SPIN, one of six centers of STARnet, a Semiconductor Research Corporation program, sponsored by MARCO and DARPA.


%

\end{document}